\documentclass{aa}
\usepackage{graphics,epsfig}
\begin{document}

\title{The globular cluster mass/low mass X-ray binary
correlation:implications for kick velocity distributions from
supernovae}

\author{Michiel Smits\inst{1}, Thomas J. Maccarone \inst{2,1}, Arunav Kundu \inst{3}, Stephen E. Zepf \inst{3}}

\offprints{T.J. Maccarone; tjm@phys.soton.ac.uk}

\institute{$^1$ Astronomical Institute 'Anton Pannekoek', University
of Amsterdam, Kruislaan 403, 1098 SJ, Amsterdam, The Netherlands\\
$^2$School of Physics and Astronomy, University of Southampton,
Southampton, Hampshire, SO17 1BJ, UK\\$^3$ Department of Physics and
Astronomy, Michigan State University, East Lansing MI, USA}

   \date{}
   \authorrunning{Smits et al.}
   \titlerunning{Globular cluster LMXBs and kicks}
   \maketitle

\def\simlt{\mathrel{\rlap{\lower 3pt\hbox{$\sim$}}
        \raise 2.0pt\hbox{$<$}}}
\def\simgt{\mathrel{\rlap{\lower 3pt\hbox{$\sim$}}
        \raise 2.0pt\hbox{$>$}}}

\label{firstpage}
\input epsf
\begin{abstract}
Optical and X-ray studies of six nearby galaxies show that the
probability a globular cluster will be an X-ray source is consistent
with being linearly proportional to its mass.  { We show that this
result is consistent with some recent estimates of the velocity kick
distributions for isolated radio pulsars -- those which are the sum of
two Maxwellians, with the slower distribution at about 100 km/sec --
so long as a large fraction of the retained binaries are in binary
systems with other massive stars.}  We confirm that over a large
sample of galaxies, metallicity is clearly a factor in determining
whether a globular cluster will contain an X-ray binary, and we
estimate the transformations between color and metallicity for a large
number of optical filter combinations.  We also show that the core
interaction rate is roughly linearly proportional to the stellar mass
of a globular cluster for the Milky Way when one bins the clusters by
mass.

\keywords{globular clusters, galaxies, X-ray binaries, supernovae}
\end{abstract}

\section{Introduction}
It has long been known that globular clusters contain much larger
numbers of X-ray binaries per unit stellar mass than do field
populations of galaxies (Katz 1975; Clark 1975).  This overabundance
has been ascribed to stellar interactions in globular clusters which
allow neutron stars and/or black holes to enter new binary systems
through either tidal capture (Clark 1975; Fabian, Pringle \& Rees
1975) or exchange interactions (Hills 1976) long after the supernovae
that produce them.  In the Chandra era, it has been possible to extend
this work to show that the clusters with the highest interaction rates
are most likely to contain X-ray sources, and probably are most likely
to contain accreting neutron stars (Pooley et al. 2003; Heinke et
al. 2003; Gendre 2005).

While studies of Galactic globular clusters have thus been fruitful,
the Milky Way's globular cluster system is rather small (containing
only about 150 clusters), so it is not possible to study certain
phenomena due to lack of statistical significance in a small sample of
clusters.  Furthermore, some of the Milky Way's globular clusters'
parameters, such as metallicity and galactocentric radius, are
strongly correlated with one another, making it difficult to isolate
the causes of certain effects.  As a result, extragalactic globular
cluster systems can be invaluable in complementing the Galactic
globular clusters for producing the best possible data on how globular
cluster parameters affect X-ray binary production.

While ROSAT was able to observe globular cluster X-ray sources in M~31
(see e.g. Supper et al. 1997), it was not until the advent of the
Chandra X-ray Observatory that there was sufficient angular resolution
to resolve apart the sources in more distant galaxies, or to obtain
sufficiently accurate positions that comparisons could be made with
optical counterparts.  Early on, by comparing the positions of X-ray
sources with those of optically detected globular clusters, it was
found that large fractions of the X-ray binaries in elliptical
galaxies were in globular clusters (Sarazin, Irwin \& Bregman 2001;
Angelini, Loewenstein \& Mushotzky 2001 -- ALM).  Larger samples of
galaxies have shown a trend where the fraction of X-ray binaries in
globular clusters in a galaxy seems to vary as a function of galaxy
type, increasing from spiral galaxies to lenticular to normal
ellipticals to cD galaxies (Maccarone, Kundu \& Zepf 2003; Juett 2005;
Irwin 2005).

From the observations of elliptical galaxies, with their large
globular cluster systems, it has been possible to find correlations
between cluster properties and the probability that a cluster would
contain an X-ray binary which are not statistically significant in the
Milky Way, or which have ambiguous interpretations in the Milky Way
because of the aforementioned correlations between different cluster
parameters.  Specifically, it has been shown clearly that metal rich
clusters are far more likely to contain X-ray binaries than metal poor
clusters (Kundu, Maccarone \& Zepf 2002), confirming suggestive
results from the Milky Way (Grindlay 1993; Bellazzini et al. 1995).
These results have been found again in numerous subsequent papers (see
e.g. Jordan et al. 2004; Minniti et al. 2004; Xu et al. 2005; Kim et
al. 2005).  It has also been shown that more luminous clusters are
more likely to contain X-ray binaries than less luminous clusters (see
e.g. ALM;KMZ).  We refer the reader also to reviews by Verbunt \&
Lewin (2006) and Fabbiano (2006) for a broader overview of the
literature.

Other results have been suggested by the data, but are not as clearly
significant.  While in the Milky Way, it is clear that denser globular
clusters are more likely to contain large numbers of X-ray sources per
unit stellar mass (Pooley et al 2003; Heinke et al 2003; Gendre 2005),
the extragalactic clusters have core radii which have not been
sufficiently well resolved spatially as to establish this effect
clearly.  KMZ found an anti-correlation between cluster half-light
radius and LMXB hosting probability in NGC~4472, but noted that it was
statistically insignificant.  Jordan et al. (2004) estimated King
model parameters of globular clusters in M87.  They then inferred
stellar collision rates ($\Gamma$) from the model fits, and found that
the collision rates were strongly correlated with the likelihood that
a cluster would contain an X-ray binary, but that this correlation was
only marginally better than the correlation with cluster luminosity.
The absence of strong signatures of cluster concentration should not
be taken as evidence that cluster concentration is uncorrelated with
X-ray binary production.  Half-light radius is not expected to be very
well correlated with core concentration, making the null result of KMZ
unsurprising.  Because the core radii of globular clusters in the
Virgo Cluster are typically a small fraction of a pixel on the
Advanced Camera for Surveys aboard the Hubble Space Telescope, neither
is it surprising that Jordan et al. (2004) were unable to find a
strong signature of cluster concentration.

In this paper, we analyze a sample of six galaxies for which there is
good optical and X-ray data.  We show unambigously that the
metallicity effect is strong.  We then isolate the effect of cluster
mass, $M$, on probability that a cluster will contain an X-ray source,
and find that the probability a globular cluster will contain an X-ray
source scales as $M^{1.03\pm0.12}$.  { We compare this value with
the expected value from various kick velocity distributions, and find
that this result is most consistent with double Gaussian distributions
for pulsar kick velocities, where there exists a slow mode to the
neutron star kick velocity distribution}.

\section{Observations}
\begin{figure}
\begin{center}
\renewcommand{\epsfsize}[2]{0.7#1}
\epsfig{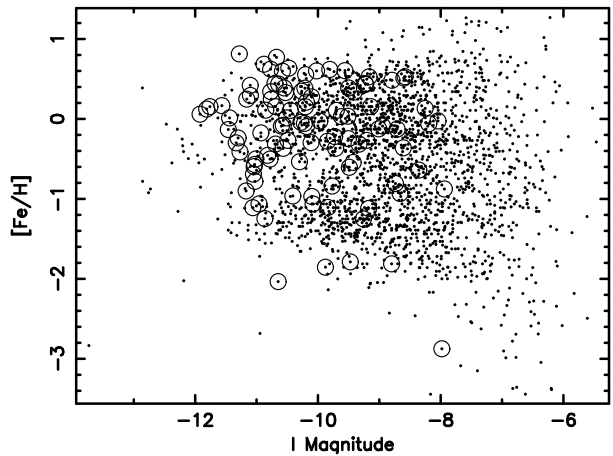}
\caption{The globular clusters analyzed in this data set.  Small dots
  represent optically detected globular clusters, while the larger open
circles surround the clusters with X-ray sources.}
\label{scatterplot}
\end{center}
\end{figure}

We use data from HST and Chandra for six galaxies: NGC~1399, NGC~3115,
NGC~3379, NGC~4472, NGC~4594, and NGC~4649.  The data analysis
procedures used here are the same as those used in our previous work
(see Maccarone et al. 2003 for details).  { We take the distances
to these galaxies from Tonry et al. (2001)}.  Partial results for all
of these galaxies have been presented previously (ALM; Kundu et
al. 2003, 2004; KMZ, Di Stefano et al. 2003), and a detailed analysis
with full source catalogs will be presented in future work (Kundu et
al. in prep.)  In all, our sample includes 98 X-ray sources in 2276
globular clusters.  These data points are plotted in Figure
\ref{scatterplot}.

Because the data are taken in different filter sets ($B-I$ for
NGC~1399 and $V-I$ for the rest of the galaxies), we must develop an
algorithm to convert from magnitudes to cluster masses and
metallicities.  The $I$-band magnitudes are used to estimate the
cluster masses, while the colors are used to estimate the cluster
metallicities.  Following the procedure of Kundu \& Whitmore (1998),
we determine the color-to-metallicity conversions for all commonly
used sets of photometric filters.  We take all globular clusters from
the Harris (1996) catalogue for which there are data in a given pair
of filters, for which there is a spectroscopic metallicity
measurement, and for which $E(B-V)<0.4$.  We then de-redden the data,
using the E(B-V) versus extinction relations from Cardelli, Clayton \&
Mathis (1989), and fit [Fe/H] as a function of color and vice versa,
and take the bisector of the two fits.  This typically produces a
color-metallicity relation with a scatter of about 0.2 dex in [Fe/H].
The fact that the scatter is typically so small (i.e. about the same
size as what would be expected from measurement errors) indicates that
linear color-metallicity relations are adequate for most purposes. The
results are given in Table 1.

\begin{table}
\caption{The color-metallicity relations for 10 different colors.}
\label{colors}
\begin{center}
\begin{tabular}{|c|c|c|}\hline
Color & N & Relation \\ \hline
V-I & 65 & [{\rm Fe/H}] = -5.79 + 4.68 (V-I)\\ \hline
U-B & 72 & [{\rm Fe/H}] = -1.91 + 3.22 (U-I)\\ \hline
B-V & 78 & [{\rm Fe/H}] = -4.83 + 4.91 (B-V)\\ \hline
V-R & 55 & [{\rm Fe/H}] = -5.93 + 9.82 (V-R)\\ \hline
U-V & 72 & [{\rm Fe/H}] = -3.15 + 2.05 (U-V)\\ \hline
B-R & 55 & [{\rm Fe/H}] = -5.49 + 3.58 (B-R)\\ \hline
B-I & 65 & [{\rm Fe/H}] = -5.87 + 2.75 (B-I)\\ \hline
U-R & 55 & [{\rm Fe/H}] = -3.61 + 1.71 (U-R)\\ \hline
U-I & 64 & [{\rm Fe/H}] = -4.19 + 1.57 (U-I)\\ \hline
R-I & 54 & [{\rm Fe/H}] = -4.41 + 6.39 (R-I)\\ \hline
\end{tabular}
\end{center}
\end{table}

\section{Analysis}
Assuming that the probability that a globular cluster contains an LMXB
can be described as a power law function of the cluster's mass times a
power law function of its metallicity, we attempted to determine the
exponents of the power laws.  First we attempted to do the fitting
using density estimation (Silverman 1986), applying Gaussian smoothing
to the two dimensional data set with different smoothing factors.  We
began by making a grid of 100$\times$100 in mass and metallicity, and
tried smoothing factors from 3 to 10 (i.e. smearing with a Gaussian
kernel with a full width half maximum ranging from 3 to 10 cells on
the grid).

While the best fitting results
always found that:
\begin{equation}
P(LMXB) \propto Z^{0.25\pm0.03},
\end{equation}
where $P(LMXB)$ is the probability a cluster will contain an X-ray
binary, and $Z$ is the cluster's metallicity.  This metallicity
exponent was found regardless of the smoothing factor, while the mass
exponent was found to depend heavily on the choice of smoothing
factor.  The reason for this is likely that near the edges of the
globular cluster distribution, gaussian smoothing is applied to a
source distribution which is asymmetric.  This tends to bias the data.
As a result, the density estimation for data sets such as ours is not
robust, { but since, as we note below, the mass exponent, which is
the focus of this paper does not depend on the metallicity exponent
assumed, we save detaied treatment of the effects of metallicity on
X-ray binary production for future work (Kundu et al., in prep.).}

The alternative is to use binning.  We then sorted the clusters by
mass, and compared the total number of X-ray binaries in
logarithmically spaced mass bins (varying the number of bins from 5 to
10) with the expression $Y$, defined as:
\begin{equation}
Y=\Sigma Z_i^{0.25} M_i,
\end{equation}
where $Z_i$ is the metallicity of a given cluster and $M_i$ is the
mass of a given cluster.  In doing this, we have assumed that the
metallicity effect does scale as [Fe/H]$^{0.25}$, although we have
found that the mass exponent is not sensitive to the choice of the
metallicity exponent.  This is as expected; mass and metallicity of
globular clusters are observed to be uncorrelated (Ashman \& Zepf
1998), so it would be suprising if the correction made for the
metallicity strongly affected the inferred correlation between mass
and cluster LMXB probability.

\begin{figure}
\begin{center}
\renewcommand{\epsfsize}[2]{0.7#1}
\epsfig{file=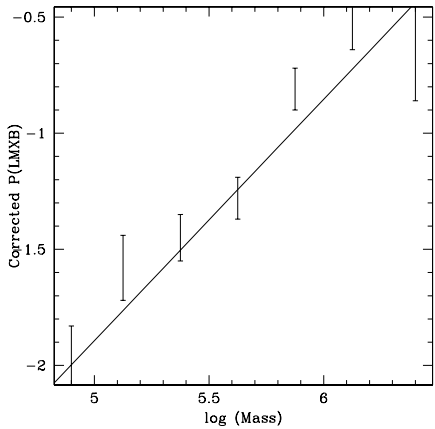,width=6cm}
\caption{Metallicity corrected LMXB probability as a function of mass,
fit using binning, with 7 bins.  Fits with different numbers of bins
produce results which are the same within statistical errors.
Changing the metallicity correction between 0.20 and 0.30 has no
effect on the mass exponent, as expected given the absence of
correlations between cluster mass and metallicity.}
\label{bin}
\end{center}
\end{figure}

We then use minimization of the Cash statistic (Cash 1979) to estimate
the mass exponent, and find that { $P({\rm LMXB})\propto
Y^{1.03\pm0.12}$ (see e.g. Figure \ref{bin} for a fit using 7 bins),
which means that $P({\rm LMXB})\propto M^{1.03\pm0.12}$, since the
exponent for $Y$ is independent of the exponent for the metallicity
term, meaning that all the dependence on $Y$ is due to its mass
dependence}.  We compare the results using different numbers of bins
and find that the results are not strongly sensitive to the number of
bins used { (for numbers of bins in mass between 5 and 10, the best
fitting value varies from 1.01 to 1.07, with a standard deviation of
0.11 or 0.12 for every fit)}. { For a given number of bins, varying
the metallicity exponent put into $Y$ changes the fitted dependence of
$P({\rm LMXB})$ on $Y$ by only one part in one thousand -- much less
than the changes due to using different numbers of bins.  We have also
made linear-logarithmic and logarithmic-linear plots of the data and
found that no reasonable fit can be made using such functions.  Thus
while a power law is obviously not a unique description of the data,
it seems to be the most reasonable simple functional form for the
data, and a reasonable choice for parameterizing the data.} We note
that for a smaller number of clusters in M~87, Jordan et al. (2004)
found a metallicity exponent of $0.3\pm0.1$ and a mass exponent of
$1.08\pm0.11$, so our results are consistent with past work.

\section{Discussion} 

\subsection{Implications of the observational results for retention fractions}
The number of X-ray binaries in a globular cluster is likely to be
correlated most strongly with the stellar collision rate.  The
collision rate should be dominated by the collisions which take place
in the core, and it is straightforward to compute analytically the
collision rate in the core of a globular cluster:
\begin{equation}
\Gamma=\rho_c^{3/2}r_c^{2},
\end{equation}
where $\rho_c$ is the central density of the cluster and $r_c$ is the
core radius.  { This formula assumes virial equilibrium in the
cluster core (Verbunt \& Hut 1987; Pooley et al. 2003), and follows
from the more general $\Gamma=\rho_c^{2}r_c^{3}/\sigma$, where
$\sigma$ is the velocity dispersion, since $\sigma\propto$$\rho_c^{0.5}
r_c$.}

\begin{figure}
\begin{center}
\renewcommand{\epsfsize}[2]{0.7#1}
\epsfig{file=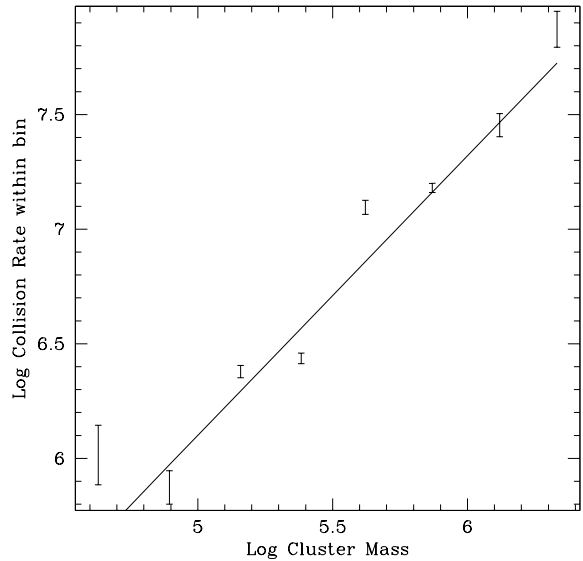,width=8cm}
\caption{Binned collision rate versus mass for the Milky Way's
globular clusters, using data from the Harris catalog with 8 bins.}
\label{bin_MW}
\end{center}
\end{figure}

The only galaxy where the core radii and central densities of a large
number of clusters have been reliably measured (rather than inferred
from model fits) is the Milky Way.  The collision rates for the Milky
Way are well correlated with the masses; when we bin the Milky Way's
globular clusters by mass and fit a power law to the data.  We find
that the best fitting power law index varies between 0.9 and 1.3,
depending on the number of bins used (see figure \ref{bin_MW} for a
characteristic fit).  { The statistical errors on the fits are always
smaller than 0.1 in index.}  We thus adopt 1.1$\pm$0.2 for the
exponent here, but note that this is a systematic error, and really
represents a hard bound on the possible values, rather than a
$1\sigma$ error. {\it This result is intermediate between the two
cases where the spatial properties are independent of mass
(i.e. where, among the King model parameters, only the central density
varies, and the expectation value would be $\Gamma\propto$$M^{1.5}$)}
and where the concentration index (i.e. the ratio of tidal radius to
core radius) is independent of mass (in which case the expectation
value would be $\Gamma\propto$$M^{2/3})$.  The intermediate result is
expected given that cluster concentration increases with increasing
cluster mass (Djorgovski \& Meylan 1994).  { It is possible to find
other means of producing various exponents besides those presented
above, if one allows for different relations between the fraction of
the mass in the cluster core and the core radius with mass.  The key
point, though, is that empirically, the relation for Milky Way
globular clusters is that $\Gamma$ scales approximately as $M^{1.1}$.}

Jordan et al. (2004) found that the probability a cluster in M~87 will
contain an X-ray binary is proportional to $\Gamma
\rho_c^{-0.4\pm0.1}$, and suggested that the deviation from a purely
linear relationship with $\Gamma$ might be due to destruction of
binaries by dynamical interactions in the globular cluster.  While
clusters do undoubtedly destroy binary systems, this should be done
preferentially for the longest period systems, perhaps by making them
highly eccentric rather than by destroying them outright if they are
Roche lobe overflowing systems(e.g. Hut \& Paczynski 1984; Rasio \&
Heggie 1995; Maccarone 2005).  Binary destruction should have
relatively little impact on the short period systems which, due to
their higher duty cycles (see e.g. Piro \& Bildsten 2002; Bildsten \&
Deloye 2004; Ivanova \& Kalogera 2006) should provide most of the {\it
bright} X-ray binaries which can be seen out to Virgo Cluster
distances (see Maccarone et al. 2005 for a discussion of the effects
of destruction of X-ray binaries due to dynamical interactions on
their orbital period distributions).

In light of our above finding that, { empirically,} for the Milky Way,
$\Gamma$ scales with cluster mass to the power 1.1$\pm$0.2, it seems
simpler just to consider that the estimates of the core radii of the
globular clusters in M~87 are likely to be dominated by measurement
uncertainties. In the absence of useful radial information, the
derived central density of a globular cluster will scale linearly with
the cluster's mass; the estimate of $\Gamma$ will then scale with the
cluster mass to the power 3/2.  We have shown that, for the Galactic
globular cluster systems, $\Gamma$ scales linearly with the cluster
mass; therefore, one will need to correct this derived value of
$\Gamma$ by a term of order $M^{1/2}$ (or $\rho^{1/2}$ in the absence,
again, or any useful radial information).  The $\rho^{-0.4}$ term
found by Jordan et al. (2004) is thus more likely this ``correction''
term than a term indicating dynamical destruction of accreting
binaries. { The only other means of producing the $\Gamma$-M
correlation we show and the relation among $\Gamma,M,$ and $\rho_c$
found by Jordan et al. (2004) would be a specific relationship between
the fraction of mass in the core and the core radius for clusters, in
which the cluster core density is uncorrelated with the cluster mass.
This would, in principle, be possible, e.g. if the most massive
clusters has progressively lower fraction of their mass in their core
but is, in fact, not the case, given the strong observed correlation
between cluster central density and cluster absolute magnitude
(Djorgovski \& Meylan 1994).}

Finally, let us consider an additional important point.  The collision
rate $\Gamma$ is a collision rate of typical stars with one another --
mostly main sequence stars with main sequence stars.  { The collision
rate for neutron stars will be the collision rate itself, times the
fraction of stars in the cluster core which are neutron stars.  This
fraction will come from { three} factors -- stellar evolution,
which determines what fraction of stellar mass ends up locked up in
neutron stars, the retention fraction, which characterizes what
fraction of those neutron stars remain in the cluster, { and the
effects of mass segregation, which determines how over-represented
neutron stars are in the cores of globular clusters, compared to their
representation in the clusters on the whole}.  We assume that the
stellar evolution effects have no mass dependence, so any difference
between the collision rate as a function of cluster mass and the LMXB
hosting probability as a function of cluster mass is most naturally
explained by the dependence of the retention fraction on cluster mass,
and mass segregation variations as a function of mass.  

{ As a result, we define a quantity, $\Phi$, which should scale
linearly with the expected formation rate of neutron star X-ray
binaries in the core of a globular cluster.  We add only a single
assumption at this point, which is that mass segregation works
sufficiently effectively on timescales of the ages of globular
clusters that all the neutron stars are in the cluster cores.  This
introduces an additional factor of $M/M_{core}$ to account for the
``overdensity'' of neutron stars among the total stars in a cluster's
core.  }Since $M_{core}\propto$$r_c^3\rho_c$, this yields:
\begin{equation}
\Phi=\Gamma M/M_{core} = \rho^{1/2}M/r_c.
\end{equation}
{ Following the same procedure as above to estimate the $\Gamma-M$
relation, we take the binned sum of the Milky Way's globular clusters'
values for these same $\Phi$.  We find that $\Phi \propto
M^{0.5-0.7}$, again with the range of estimated exponent values
dominated by the binning scheme used (i.e. whether 6, 8 or 10 bins are
used) rather than by measurement uncertainties (see e.g. Figure
{\ref{phiplot}).  Since $\Phi f_{ret}$ should give the LMXB hosting
probability, that leads to the inference that $f_{ret} \propto
M^{0.4-0.6}$, since the hosting probability scales as $M^{1.1}$.

We also note that mass segregation of neutron stars is not
sufficiently fast as to place {\it all} neutron stars in the cores of
globular clusters; there are some millisecond pulsars with
well-measured positions that place them outside the cores of their
host clusters (see e.g. Camilo \& Rasio 2005 for a reasonably
up-to-date list of globular cluster pulsars, or Paulo Freire's
continuously updated list at http://www.naic.edu/~pfreire/GCpsr.html).
While, relaxing this assumption leads to a weaker dependence of the
retention fraction on mass, the assumption is generally a good one.
About half of all misllisecond pulsars are within one core radius of
the center of their host clusters, while the fraction of the total
mass in the core is less than 10\% even in 47~Tuc, which has the
highest core fraction of any massive cluster.}

\begin{figure}
\begin{center}
\renewcommand{\epsfsize}[2]{0.7#1}
\epsfig{file=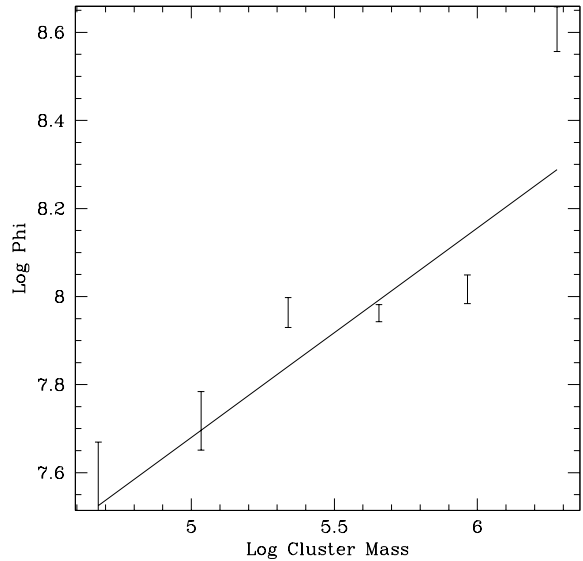,width=8cm}
\caption{Binned values of $\Phi$ versus mass for the Milky Way's
globular clusters, using data from the Harris catalog with 6 bins.}
\label{phiplot}
\end{center}
\end{figure}

{\subsection{Scaling relations of the retention fraction}} We have
considered a variety of possible kick velocity distributions in order
to determine whether any of them { are consistent with the
estimated retention fraction relation -- $F_{ret}\propto M^{0.5}$}.
Essentially, two broad classes of velocity kick distributions have
been proposed for radio pulsars: ``zero-peaked'' distributions
(e.g. Paczynski 1990; Hartman 1997; Hansen \& Phinney 1997), where the
most likely initial velocity for the neutron star is close to zero
(although the distribution retains significant probability density out
to hundreds of km/sec), and distributions which are Gaussians peaked
at the mean pulsar velocity (e.g. Lyne \& Lorimer 1994) or
Maxwellians, which have rather similar characteristics.  Some recent
work has modelled the pulsars' velocity distributions as double
Gaussians { or double Maxwellians} (Arzoumanian, Chernoff \& Cordes
2002; Brisken et al. 2003), typically with one peak at about 100
km/sec and containing about 20\% of the neutron stars and the other
peak at about 300-500 km/sec containing about 80\% of the neutron
stars.  In essence, the two approaches have converged to rather
similar distributions, since the low velocity { mode} ``fills in''
the gap left at low velocities in the single { mode} approach.  On
the other hand, the most recent estimation of the kick velocity
distribution argued for a return to a single-peaked Maxwellian
distribution with a characteristic { 1D velocity of about 265
km/sec} (Hobbs et al. 2005), so there still exists controversy about
the initial velocity distribution of radio pulsars.}

{ \subsubsection{Results of models tried} We have tried a variety of
different models with different parameter values to determine whether
they can produce retention fractions with negligible mass dependence.
We present the methodology and a sampling of the results here.  In the
interests of brevity, we do not present every calculation we have
done, but note that these are in the Master's degree thesis of the
lead author of this paper { albeit with a somewhat different
interpretation section} and can be obtained by contacting the authors
of this paper.  

To calculate reliable retention fractions we use 111 globular clusters
from the Harris catalogue as templates. We used the core and tidal
radii from the catalogue, converting arcminutes to parsecs using the
distances provided in the catalogue. We also converted the V magnitude
to mass assuming a constant mass-to-light ratio of 1.8, as found by
Piatek et al. (1994). The 111 clusters are all clusters from the
catalogue that are not core-collapsed and include the 3 parameters we
need for fitting.

We use two different models to describe the early globular
clusters. The first is the King model fitted to present-day average GC
properties (from the Harris catalogue). All heavy stars are assumed to
be in the center of the cluster at the end of their lifetimes and the
retention fractions are calculated using the central escape
velocity. The second model assumes that no mass segregation has taken
place. Stars of all masses are spread equally through the cluster
according to the cluster's mass density profile. The Plummer model is
used because it puts the GCs with different central potentials on
equal footing. GCs evolve to small core radii and large tidal radii as
they age. This means that the present-day central potential has no
meaning when applied to primordial GCs. Through dynamical friction,
the high mass stars may fall to the center before their maximum age is
reached or they may end somewhere between where they started and the
center of the cluster.  The escape velocity needed for the stellar
remnant is estimated accordingly.  In each cases, we simulate the
results (i.e. retention or ejection) for 10000 supernova explosions
producing neutron stars.

We use the equations from Drukier(1996) to calculate retention
fractions. The probability that a star is retained to the cluster
after applying the supernova "kick" is as follows:\\
\begin{equation}
\label{retention}
P(ret|\hat{v}, \hat{v_{k}}) = 
\left\{ 
\begin{array}{ll} 1 & \hat{v_{k}} -1 \leq -\hat{v};\\ \frac{1 - (\hat{v} -\hat{v_{k}})^{2}}{4\hat{v}\hat{v_{k}}} & |\hat{v_{k}} -1| < \hat{v};\\ 0 & \hat{v_{k}} -1 \geq \hat{v}
\end{array} 
\right\},
\end{equation}
where $\hat{v_{k}} = \frac{v_{k}}{v_{e}}$ and $\hat{v} = \frac{v}{v_{e}}$, meaning both the pre-kick velocity and the kick
velocity are divided by the escape velocity. $\hat{v}$ is taken to be
a Maxwellian distribution between 0 and 1. Any $\hat{v} > 1$ would
escape from the cluster, therefore we use a lowered Maxwellian
distribution.  The escape velocities are taken from our Plummer and
King models. We use the central escape velocity for both models, but
we also use a Plummer model with no primordial mass segregation as
explained in the previous subsection. The King model escape velocities
are the highest (typically 25 km/s), because most of the GC cores are
more concentrated than the cores of the Plummer models (typically 15
km/s). The model with no primordial mass segregation has the lowest
escape velocities (typically 8 km/s), because the supernova
progenitors are not concentrated in the center of the core.

As a simple way to test the method we considered a constant kick
velocity distribution. We tried 10 progressively larger kick velocity
intervals. The intervals we used are [0,x] km, with x = 500, 600,
...., 1400. We use the 3 different models: King, P-center (Plummer
model, all heavy stars in the core) and P-spread (Plummer model with
the stars spread through the GC, no primordial mass segregation) in
most cases, although since the different models result in
qualitatively similar conclusions, we do not exhaustively present all
the results. To each we fit a power law for the retention fraction as
a function of mass (figure \ref{ret}).  The plots for all other kick
velocity distributions are qualitatively similar, so for the remainder
of the paper, we present tables showing the results rather than plots.
We obtain the exponent and the average retention fraction. The results
are in table \ref{constant}. The exponents are all approximately
0.50. This is because the escape velocity goes as the square-root of
the cluster mass. The King model has a slightly higher exponent
because more massive clusters tend to favor a slightly higher central
potential and are more concentrated, resulting in a somewhat higher
escape velocity. The average retention fraction is linear with x. This
is as expected. If a constant probability distribution is spread out
over an interval twice as large (0-1000 instead of 0-500), one expects
half as many neutron stars to be retained.

\begin{figure}
\begin{center}
\renewcommand{\epsfsize}[2]{0.7#1}
\epsfig{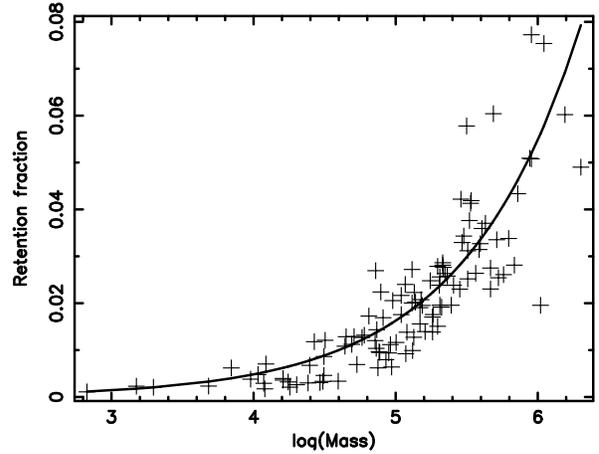}
\caption{The retention fraction as a function of mass, fitted to the King-model data with a constant kick velocity distribution between 0 and 1000 km/s. Chi-square was minimized to obtain the fit.}
\label{ret}
\end{center}
\end{figure}

\begin{table*}[!h]
\label{constant}
\caption{Retention fraction fit results for a constant kick velocity distribution}
\begin{center}
\begin{tabular}{|c|c|c|c|c|c|c|} \hline
& \multicolumn{2}{c|}{King} & \multicolumn{2}{c|}{P-center} & \multicolumn{2}{c|}{P-spread} \\\hline
x &exponent&fraction&exponent&fraction&exponent&fraction\\\hline
500 & 0.53 & 0.041 & 0.50 & 0.018 & 0.50 & 0.0053 \\\hline
600 & 0.53 & 0.034 & 0.50 & 0.015 & 0.50 & 0.0044 \\\hline
700 & 0.53 & 0.029 & 0.50 & 0.013 & 0.50 & 0.0038 \\\hline
800 & 0.53 & 0.026 & 0.50 & 0.011 & 0.50 & 0.0033 \\\hline
900 & 0.53 & 0.023 & 0.50 & 0.010 & 0.50 & 0.0029 \\\hline
1000 & 0.53 & 0.021 & 0.50 & 0.009 & 0.50 & 0.0026 \\\hline
1100 & 0.53 & 0.019 & 0.50 & 0.008 & 0.50 & 0.0024 \\\hline
1200 & 0.53 & 0.017 & 0.50 & 0.008 & 0.51 & 0.0022 \\\hline
1300 & 0.53 & 0.016 & 0.50 & 0.007 & 0.51 & 0.0020 \\\hline
1400 & 0.53 & 0.015 & 0.50 & 0.006 & 0.51 & 0.0019 \\\hline
\end{tabular}
\end{center}
\end{table*}
We have made similar studies of the effects of linear kick
distributions (i.e. where the probability distribution varies
linearly, with a maximum at $v=0$, and a probability of zero above
some characteristic velocity) and Gaussian velocity kick distributions
(which are taken to be Gaussian probability density functions with the
characteristic velocity equal to the spread in the velocity
distribution).  The linear distributions are characterized in terms of
a characteristic velocity $x$, and are presented for large and small
velocities in tables \ref{linear} and \ref{linear2} respectively.  The
linear distributions, like the constant probability density
distributions, generically produce a $M^{0.5}$ retention fraction mass
dependence, unless the characteristic velocity is less than the escape
velocity of the most massive clusters, in which case the dependence
will be flatter.  Both these functional forms are thus consistent with
the observations of extragalactic X-ray binaries, but will most likely
have trouble explaining the dearth of very low velocity pulsars, and
simultaneously explaining the small, but non-neglible numbers of very
high velocity pulsars (e.g. those with velocities of more than 800
km/sec) while still producing retention fractions greater than $\sim$
1\%, which is needed to explain the large numbers of X-ray binaries
and millisecond pulsars in globular clusters.

Gaussian distributions produce a wide range of mass dependences, since
for a non-zero Gaussian, there typically will exist a positive slope
in the probability density distribution over the range of globular
cluster escape velocities.  The results of Gaussian distribution
studies are shown in table \ref{Gaussian}.  It is clear that to
reproduce the observations, a Gaussian distribution would need a
characteristic velocity just less than 20 km/sec.  This is, on its
face, strongly inconsistent with the proposed existing kick velocity
distributions, but, upon consideration of the effects of binaries, it
may become consistent with e.g. the results of Arzoumanian et
al. (2002).  The presence of a heavy binary companion at the time of
the supernova explosion will dilute the velocity kick applied to the
system, making retention more likely; this has already been suggested
to be a big part of the solution to the retention fraction problem
(e.g. Davies \& Hansen 1998; { Pfahl et al. 2002}).  The maximum
possible dilution would occur if the envelope of the proto-neutron
star has been completely stripped in the binary, while the companion
star is very close to the maximum possible mass in order not to
undergo a supernova explosion itself later.  That is to say, the
maximum dilution occurs for a supernova explosion of a 1.4 $M_\odot$
core with an 8 $M_\odot$ companion, and this dilution will be a factor
of 6.7 [i.e. 1.4/(1.4+8.0)].  This would be enough to bring a 90
km/sec kick (the characteristic kick velocity of the slow mode found
by Arzoumanian et al. 2002) down to about 13 km/sec {(we do note
that the Arzoumanian model is a Maxwellian, rather than a Gaussian,
and a 90 km/sec Maxwellian will have slightly few low velocity objects
than a 90 km/sec Gaussian, but that this should lead to only a minor
difference)}.  Thus we can say that the most extreme dilution factor
is more than sufficient to produce a reasonable dependence of the
retention fraction on mass.  Additionally, heavy binaries will have
lower initial velocities than the stars in the cluster as a whole,
which should allow them to accomodate larger kicks (although in no
case will this increase the kick velocity allowed by more than
$\sqrt{2}$).  Since mass segregation will bring the most massive stars
to the centers of clusters, and these massive stars will have the
highest cross-sections for interactions, it is likely that supernova
progenitors will find themselves in binaries with other massive stars.
Whether this actually happens to the extent required by the data must
be addressed through numerical simulations and is beyond the scope of
this paper.

Regardless, it seems that in no case will the data be consistent with
a single Maxwellian or Gaussian with a post-dilution characteristic
velocity greater than 20 km/sec.  This seems to rule out the Hobbs et
al. (2005) suggestion that the pulsar velocity kick distribution can
be a single Maxwellian at 265 km/sec, unless the slower kicks in
globular clusters are related to binary evolution (see e.g. Pfahl et
al. 2002b; Dewi et al. 2005 for suggestions that binary evolution may
affect pulsar kick velocities).  It also indicates that the
contribution of neutron stars from the fast mode in a distribution
like that of Arzoumanian et al. (2002) to retention fractions should
be negligible.  { These results are thus quite similar to the
conclusions of Pfahl et al. (2002) who found that, while a single fast
kick mode had trouble producing neutron star retention fractions large
enough to match the observed numbers of millisecond pulsars in
globular clusters like 47~Tuc, including also a slow velocity kick
mode, and the effects of binaries could potentially solve the
retention problem.}  With careful numerical simulations, the globular
cluster X-ray binaries are likely to present the strongest constraints
on the low velocity end of the pulsar kick velocity distribution,
since field studies will always be complicated, e.g., by the intrinsic
velocity dispersion in the field.

To obtain strong constraints on the actual asymmetric kick velocity
applied to neutron stars at the times of their birth will require
numerical simulations which can take account of both the distributions
of the neutron star progenitors in the cluster at the time the
supernovae occur, and what fraction of the neutron stars are in
binaries, and what are the masses of those binary companions.  The
numerical simulations will also need to be made for ranges of cluster
masses and initial concentration indices.  At the present, the best
studies of neutron star retention (e.g. Drukier 1996; Davies \& Hansen
1998; Pfahl, Rappaport \& Podsiadlowski 2002) have each considered
several aspects of this problem; the simulation closest to meeting
these requirements was that of Pfahl et al. (2002), in which all the
relevant factors except for a range in cluster escape velocities were
considered.

\begin{table*}[!h]
\label{linear}
\caption{Retention fraction fit results for a linear kick velocity distribution with negative slope}
\begin{center}
\begin{tabular}{|c|c|c|c|c|c|} \hline
& \multicolumn{2}{c|}{King} & \multicolumn{2}{c|}{P-center}\\\hline
x & exponent&fraction&exponent&fraction\\\hline
100 & 0.44 & 0.348 & 0.47 & 0.168 \\\hline
200 & 0.49 & 0.187 & 0.48 & 0.087 \\\hline
300 & 0.50 & 0.129 & 0.49 & 0.059 \\\hline
400 & 0.51 & 0.098 & 0.49 & 0.044 \\\hline
500 & 0.51 & 0.080 & 0.49 & 0.036 \\\hline
600 & 0.51 & 0.066 & 0.49 & 0.030 \\\hline
700 & 0.51 & 0.057 & 0.49 & 0.026 \\\hline
800 & 0.52 & 0.050 & 0.49 & 0.022 \\\hline
900 & 0.52 & 0.045 & 0.49 & 0.020 \\\hline
1000 & 0.52 & 0.040 & 0.49 & 0.018\\\hline
\end{tabular}
\end{center}
\end{table*}

\begin{table}[!h]
\label{linear2}
\caption{Retention fraction fit results for a linear kick velocity distribution with very large negative slope, for a King model}
\begin{center}
\begin{tabular}{|c|c|c|} \hline
& \multicolumn{2}{c|}{King}\\\hline
x & exponent&fraction\\\hline
10 & 0.11 & 0.90\\\hline
20 & 0.20 & 0.79\\\hline
30 & 0.27 & 0.70\\\hline
40 & 0.31 & 0.61\\\hline
50 & 0.35 & 0.54\\\hline
60 & 0.38 & 0.49\\\hline
70 & 0.40 & 0.44\\\hline
80 & 0.42 & 0.40\\\hline
90 & 0.43 & 0.37\\\hline
100 & 0.44 & 0.34\\\hline
\end{tabular}
\end{center}
\end{table}

\begin{table*}[!h]
\label{Gaussian}
\caption{Retention fraction fit results for a Gaussian kick velocity
distribution.  The exponent values with asterisks denote cases where
the retention fractions are quite small and the estimations suffer
from small number statistics.}
\begin{center}
\begin{tabular}{|c|c|c|c|c|c|c|} \hline
& \multicolumn{2}{c|}{King} & \multicolumn{2}{c|}{P-center} & \multicolumn{2}{c|}{P-spread} \\\hline
$\sigma$ & exponent&fraction&exponent&fraction&exponent&fraction\\\hline
20 & 0.68 & 0.260 & 1.15 & 0.06 & 1.29 & 0.0058 \\\hline
40 & 1.06 & 0.051 & 1.43 & 0.005 & 1.46 & 0.0008 \\\hline
60 & 1.18 & 0.017 & 1.52 & 0.001 & 1.57 & 0.0002 \\\hline
80 & 1.23 & 0.007 & 1.60* & 0.000 & 1.66* & 0.0001 \\\hline
100 & 1.26 & 0.003 & 1.69* & 0.000 & 1.77* & 0.0000 \\\hline
120 & 1.28 & 0.002 & 1.81* & 0.000 & 1.89* & 0.0000 \\\hline
140 & 1.30 & 0.001 & 1.94* & 0.000 & 2.01* & 0.0000 \\\hline
160 & 1.30* & 0.001 & 2.09* & 0.000 & 2.13* & 0.0000 \\\hline
180 & 1.32* & 0.001 & 2.25* & 0.000 & 2.26* & 0.0000 \\\hline
200 & 1.33* & 0.000 & 2.42* & 0.000 & 2.38* & 0.0000 \\\hline
\end{tabular}
\end{center}
\end{table*}
}

\begin{acknowledgements}
We are grateful to Michiel van der Klis, Ralph Wijers, Simon Portegies
Zwart, Christian Knigge and Rudy Wijnands for useful discussions.  We
thank Fred Rasio for pointing out that collision rates should be
computed for neutron stars, rather than for main sequence stars, when
attempting to calculate the X-ray binary production rate.  We are
indebted to Alessia Gualandris for a variety of useful comments and
discussions.  We thank the anonymous referee for encouraging us to
discuss the observational results and theoretical calculations in more
detail and to clarify several issues in the paper{ and for
additional suggestions which dramatically improved the quality of the
paper.}  MS notes that a previous version of this work has been
submitted as a master's thesis to the University of Amsterdam.  AK and
SEZ acknowledge support from Chandra grant AR-6013X and NASA LTSA
grants NAG5-12975 and NAG5-11319.
\end{acknowledgements}

\label{lastpage}

\begin{thebibliography}{}
\bibitem{}Ashman K.M., Zepf S.E., 1998, {it Globular Cluster Systems},
Cambridge University Press
\bibitem{}Angelini L., Loewenstein M., Mushotzky R.F., 2001, ApJL, 557, 35
\bibitem{}Arzoumanian Z., Chernoff D.F., Cordes J.M., 2002, ApJ, 568, 289
\bibitem[]{480}Bellazzini, M.,Pasquali, A., Federici, L., Ferraro, F. R.,
Pecci, F. Fusi, 1995, ApJ, 439, 687
\bibitem{}Bildsten L., Deloye C.J., 2004, ApJ, 607L, 119
\bibitem{}Brisken W.F., Fruchter A.S., Goss W.M., Herrnstein R.M., Thorsett S.E., 2003, AJ, 126, 3090
\bibitem{}Cardelli J.A., Clayton G.C., Mathis J.S., 1989, ApJ, 345, 245
\bibitem[]{484}Clark, G.W., 1975, ApJL, 199, 143
\bibitem{}Davies M.B., Hansen B.M.S., 1998, MNRAS, 301, 15
\bibitem{}Dewi J.D.M., Podsiadlowski P., Pols O.R., 2005, MNRAS, 363L, 71
\bibitem{}Di Stefano R., Kong  A. K. H., Van Dalfsen  M. L., Harris W. E., Murray S. S., Delain K. M.,  2003, ApJ, 599, 1067
\bibitem{}Djorgovski S., Meylan G., 1994, AJ, 108, 1292
\bibitem{}Drukier G., 1996, MNRAS, 280, 498
\bibitem{}Fabbiano G., 2006, ARA\&A, in press, astro-ph/051148
\bibitem[]{492}Fabian, A.C., Pringle, J.E. \& Rees, M.J., 1975, MNRAS, 172, 15P
\bibitem{}Fryer C., Kalogera V., 2001, ApJ, 554, 548
\bibitem[]{495}Grindlay, J.E., 1993, ASP Conference Series 48, 156
\bibitem{}Gendre B., 2005, A\&A, 433, 137
\bibitem{}Hansen B.M.S., Phinney E.S., 1997, MNRAS, 291, 569
\bibitem{}Harris W.E., 1996, AJ, 112, 1487
\bibitem{}Hartman J.W., 1997, A\&A, 322,127
\bibitem{b10} Heinke, C.O., Grindlay, J.E., Lugger, P.M., Cohn, H.N.,
Edmonds, P.D., Lloyd, D.A. \& Cool, A.L., 2003, ApJ, 598, 501
\bibitem[]{500} Hills, J.G., 1976, MNRAS, 175, 1MNRAS, 360, 974
\bibitem{}Hobbs G., Lorimer D.R., Lyne A.G., Kramer M., 2005, MNRAS, 360, 974
\bibitem{}Hut P., Paczynski B., 1984, ApJ, 284, 675
\bibitem{}Irwin J.A., 2005, ApJ, 631, 511
\bibitem{}Ivanova N., Kalogera V., 2006, ApJ, 636, 985
\bibitem{}Jordan A., et al., 2004, ApJ, 613, 279
\bibitem{}Juett A., 2005, ApJL, 621, 25
\bibitem{}Kalogera V., King A.R., Rasio F.A., 2004, ApJ, 601L, 171
\bibitem{}Katz J.I., 1975, Nature, 253, 698
\bibitem{}Kim E., Kim D.-W., Fabbiano G., Lee M.G., Park H.S., Geisler D., Dirsch B., 2006, submitted to ApJ, astro-ph/0510817
\bibitem{b3}Kundu, A., Maccarone, T.J. \& Zepf, S.E., 2002, ApJL, 574,
5 (KMZ02)
\bibitem{b4}Kundu, A., Maccarone, T.J., Zepf, S.E. \& Puzia, T.H.,
2003, ApJL, 589, 81
\bibitem[]{509}Kundu, A. \& Whitmore, B.C., 1998, AJ, 116, 2841
\bibitem{}Lyne A.G., Lorimer D.R., 1994, Nature, 369, 127
\bibitem{}Maccarone T.J., 2005, MNRAS, 364, 971
\bibitem{b5}Maccarone, T.J., Kundu, A. \& Zepf, S.E., 2003, ApJ, 586,
814 (MKZ03)
\bibitem{}Minniti D., Rejkuba M., Funes J.G., Akiyama S., 2004, ApJ, 600, 716
\bibitem{}Mirabel I.F., Rodrigues I., 2003, Science, 300, 1119
\bibitem{}Paczynski B., 1990, ApJ, 348, 485
\bibitem{}Pfahl E., Rappaport S., Podsiadlowski P., Spruit H., 2002b, ApJ, 574, 364
\bibitem{}Pfahl E., Rappaport S., Podsiadlowski P., 2002, ApJ, 573, 283
\bibitem{}Piatek S., Pryor C., McClure R.D., Fletcher J.M., Hesser J.E., 1994, AJ, 107, 1397 
\bibitem{b13}Pooley, D. et al., 2003, ApJL, 591, 131L
\bibitem{}Portegies Zwart S.F., McMillan S.L.W., 2000, ApJL, 528, 17
\bibitem{}Rasio F.A., Heggie D.C., 1995, ApJL, 445, 133
\bibitem{}Sarazin C.L., Irwin J.A., Bregman J.N., 2001, ApJ, 556, 533
\bibitem{}Silverman B.W., 1986, {\it Density Estimation}, eds. D.R. Cox, D.V. hinkley, D. Rubin, Chapman \& Hall Ltd:London 
\bibitem{}Supper R., Hasinger G., Lewin W.H.G., Magnier E.A., van Paradijs J., Pietsch W., Read A.M., Trumper J., 2001, A\&A, 363, 63
\bibitem{}Tonry J.L., Dressler A., Blakeslee J.P., Ajhar E.A., Fletcher A.B., Luppino G.A., Metzger M.R., Moore C.B., 2001, ApJ, 546, 681 
\bibitem{}Xu Y., Xu H., Zhang Z., Kundu A., Wang Y., Wu X.-P., 2005 ApJ, 631, 809
\bibitem{}Verbunt F., Hut P., 1987, IAUS, 125, 187
\bibitem{}Verbunt F., Lewin W.H.G., 2006, to appear as Chapter 8 in "Compact Stellar X-ray Sources", eds. W.H.G. Lewin and M. van der Klis, Cambridge University Press (astro-ph/0404136)
\end{thebibliography}
\end{document}